\documentclass[pre,aps,twocolumn,amsmath,amssymb,showpacs,showkeys,unsortedaddress]{revtex4-1}
\usepackage{multirow}
\usepackage{graphicx}
\usepackage{psfrag}
\usepackage{epsfig}
\usepackage{dcolumn}
\usepackage{bm}

\newcommand{\eq}[1]{Eq.~(\ref{#1})}    
\newcommand{\fg}[1]{Fig.~\ref{#1}}     

\begin{document}
\title{A threshold induced phase transition in the kinetic exchange models}

\author{Asim Ghosh}
 \email{asim.ghosh@saha.ac.in}
 \affiliation{Theoretical Condensed Matter Physics Division, \\ Saha Institute of Nuclear Physics,
1/AF Bidhannagar, Kolkata 700 064, India}

\author{Urna Basu}
 \email{urna.basu@saha.ac.in}
 \affiliation{Theoretical Condensed Matter Physics Division,\\ Saha Institute of Nuclear Physics,
1/AF Bidhannagar, Kolkata 700 064, India}

\author{Anirban Chakraborti}
  \email{anirban.chakraborti@ecp.fr}
   \affiliation{Chaire de Finance Quantitative, Laboratoire de Math\'ematiques Appliqu\'ees aux Syst\`emes, \'Ecole Centrale Paris, 92290 Ch\^atenay-Malabry, France}
   
\author{Bikas K. Chakrabarti}
 \email{bikask.chakrabarti@saha.ac.in}
 \affiliation{Theoretical Condensed Matter Physics Division \\ \& Centre for Applied Mathematics and Computational Science,
Saha Institute of Nuclear Physics, 1/AF Bidhannagar, Kolkata 700 064, India}
\affiliation{Economic Research Unit, Indian Statistical Institute, 203 B. T. Road, Kolkata 700 108, India}

\date{\today}

\begin{abstract}
We study an ideal-gas-like model  where the particles exchange energy stochastically, through energy conserving scattering processes, which take place  \textit{if and only if} at least one of the two particles has energy  below a certain energy threshold (interactions are initiated by such low energy particles). This model has an intriguing phase transition in the sense that there is a critical value of the  energy threshold, below which the  number of particles  in the steady state goes to  zero, and above which the average number of particles in the steady state is non-zero. This  phase transition is associated with standard features like ``critical slowing down'' and  non-trivial values of some critical exponents characterizing the variation of thermodynamic quantities near the threshold energy. The features are exhibited not only in the mean field version but also in the lattice versions. 
\end{abstract}


\maketitle

\section{Introduction}
\label{intro}
The kinetic theory of gases had played a pivotal role in the development of statistical mechanics, which is more than a century old. This theory describes a gas as a collection of a large number of particles (atoms or molecules) which are constantly in random motion, and these rapidly moving particles constantly collide with each other and exchange energy. In the ideal gas, this energy is only kinetic. Recently physicists have been studying two-body kinetic exchange models in the socio-economic contexts in the  rapidly growing interdisciplinary field of ``Sociophysics'' \cite{CCCbook2006} and ``Econophysics'' \cite{SCCCbook2010}. The two-body exchange dynamics has been developed in the context of modeling income, money or wealth distributions in a society \cite{Yakovenko2009,Arnab2007,Patriarca2010,Chakraborti2010a,Redner2010,Chatterjee2004a,Chakraborti2000a,Chakraborti2009}, and modeling opinion formation in the society \cite{Mehdi2010,LCCC2010,CC2010,Sen}, analogous to the kinetic theory model of ideal gases. These studies have given deeper insights and different perspectives in the simple physics of two body kinetic exchange dynamics. In this context of wealth exchange processes, Iglesias et al. \cite{Iglesias-physica, Iglesias-sc&cul} had considered a model for the economy where the poorest in the society (atom with least energy in the gas)  at any stage takes the initiative to go for a trade (random wealth / energy exchange) with anyone else. Interestingly, in the steady state, one obtained a self-organized poverty line, below which none could be found and above which, a standard exponential decay of the distribution (Gibbs) was obtained.

Here, we study a model where  $N$ particles, interact among themselves through two-body energy ($x$) conserving stochastic scatterings with at least one of the particles having energy below a threshold $\theta$ (poverty line in the equivalent economic model). The states of particles are characterized by the energy $\{x_i\},~i=1,2,\dots,N$, such that $x_i >0, \quad \forall i$ and the total energy $E=\sum_{i} x_i$ is conserved (= $N$ here, such that the  average energy of the system $ \bar{E} = E/N = 1$). The evolution of the system is carried out according to the following dynamics:
\begin{eqnarray}
  {x^<_i}'  &=& \epsilon (x^<_i + x_j) \, ,
  \nonumber \\
 x_j' &=& (1-\epsilon) (x^<_i + x_j) \, ,
  \label{eq:ed1}
\end{eqnarray}
where $x_i^<<\theta$ (threshold energy or  `` poverty line'') and $\epsilon$ $(0 \le \epsilon \le 1)$ is a stochastic variable, changing with time (scattering). It can be noticed that, the quantity $x$ is conserved
during each collision: ${x^<_i}'+x_j' = x^<_i + x_j$. The question of interest is: ``What is the steady state distribution  $p(x)$ of energy $x$  in such systems?''

In the standard case, when the threshold energy goes to infinity ($\theta \rightarrow \infty$), we know that the steady state energy distribution will be the exponential Gibbs distribution ($p(x)\sim\exp{(-x)}$) \cite{Yakovenko2009}. However, when a finite threshold energy is introduced ($\theta > 0$), several  new and intriguing features appear. These features are exhibited not only in the mean field version (with infinite-range interaction, pairs of particles randomly chosen from $N$ particles) but also in the lattice versions (with nearest neighbor interactions, i.e. exchanges between the nearest neighbors on lattice sites). 

\section{Model simulations and results}
\label{body}

\subsection{The Model}
\label{model}
We simulate a system of  $N$ particles (agents). At any time $t$, we select randomly a particle $i$. If the energy of the particle is below a prescribed threshold energy $\theta$, then it collides with any other random particle $j$ (in the mean field model) which can have any energy whatsoever, and the two particles will exchange energy according to the Gibbs-Boltzmann dynamics of \eq{eq:ed1}. After each such successful collision, the time is incremented by unity. The dynamics will continue for an indefinite period, unless there is no particle left below the threshold energy, in which case the dynamics will freeze. If the dynamics gets frozen (when $x_i>\theta$ for all $i$),  we employ a `mild' perturbation such that a randomly chosen particle will be dropped  to the lower level ($< \theta$)  by giving up its energy to anyone else (to ensure total energy conservation).   
It can be shown that the addition of this perturbation does not alter the relevant quantities for a thermodynamically large systems, and simply ensures ergodicity in the system.
After sufficiently large  time $t> \tau$, a steady state is reached when the energy distribution    $p(x)$ (and also other average quantities) do not change with time. We start with different initial random configurations, where the states of particles are characterized by the energies $\{x_i\},~i=1,2,\dots,N$, which are drawn randomly from an uniform distribution such that $x_i >0, \quad \forall i$ and the average energy $\bar{E}=\sum_{i} x_i/N$ is set to unity. We find the system to be \textit{ergodic} (the steady state distribution $p(x)$ is \textit{independent} of the initial conditions $\{x_i\}$), and  we take steady state   averages over all such independent initial conditions to evaluate the quantities of interest.

We study mainly three cases (a) mean field (or infinite range) case where $i$ and $j$ in \eq{eq:ed1} can represent any two particles/agents in the system; (b) one dimensional case where $j = i\pm 1$ along a chain and (c) two dimensional case, where $j = i\pm \delta$ where $\delta$ represents neighbors of $i$. In our studies we consider a 2D-square lattice.

We observe  that for finite values of the energy threshold $\theta$, the steady state energy distribution is no longer the simple Gibbs-Boltzmann distribution. We also find that $O$ ($ \equiv \int_{0}^{\theta} p(x) dx$), the average number of particles below the threshold energy in the steady state, is zero for $\theta$ values below or at a  critical threshold energy $\theta_c$, and  for $\theta >\theta_c$, $O$ is non-zero.  The steady state value of $O$, the average number of particles below the threshold energy $\theta$ is seen to act like an ``order parameter'' of the system. We study the relaxation dynamics in the system: the relaxation of $O(t)$ to the steady state value of $O$ ($= O(\theta)$ for $t>\tau (\theta)$, the ``relaxation time''). We find $\tau (\theta)$ grows as $\theta$ approaches $\theta_c$, and eventually diverges at $\theta_c$. The details of the results are given below.

%

\begin{figure}
\begin{center}
    \includegraphics[angle=0,width=1.0\linewidth]
        {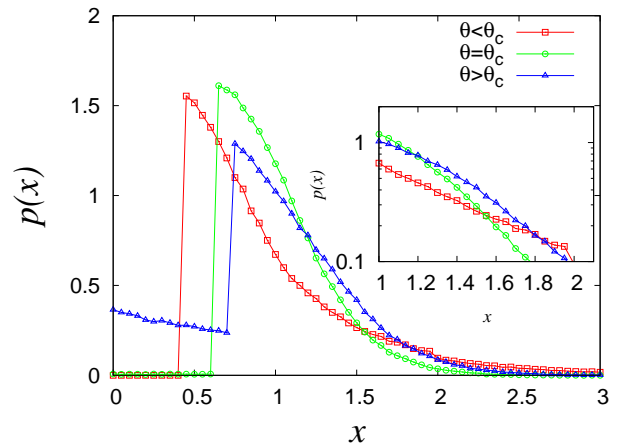}
    \end{center}
\caption{(Color online) Energy  distribution  $p(x)$ in the steady state ($t>\tau$), for different $\theta$ values.
Th inset shows semi-log plot of the energy distribution. The tail of the distribution is Gibbs-like ($N=10^5$; Mean field model with average taken over many independent initial conditions).
}
\label{fig:mf_dist}
\end{figure}

\begin{figure} [h]
\begin{center}
    \includegraphics[angle=0,width=1.0\linewidth]
        {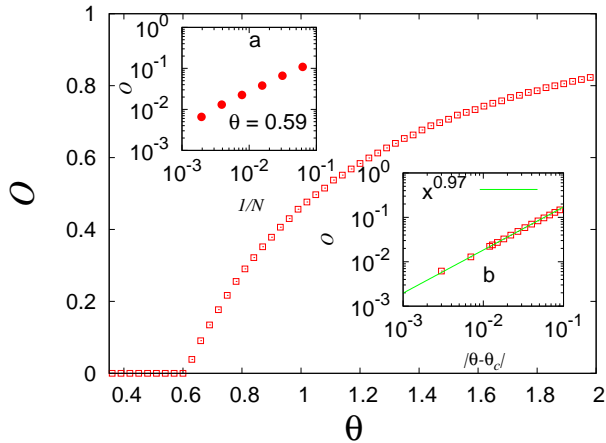}
    \end{center}
\caption{(Color online) Simulation results for the variation of $O$, the average number of particles  below the threshold energy $\theta$  in the steady state ($t> \tau$), against threshold energy $\theta$. (Inset)(a) Shows the results at $O\to 0$ for $\theta = 0.59 (<\theta_c)$ as $N\to \infty$; (b) Shows scaling fit $(\theta-\theta_c)^{\beta}$ with $\beta\simeq 0.97$. ($N=10^5$; Mean field model with average taken over many independent initial conditions).
}
\label{fig:mf_order}
\end{figure}

\subsection{Results: Mean field model}
\label{analyses}

In the mean field (long range) model, we first look for any particle ($i$) with energy $x_i<\theta$ and then this particle is allowed  to interact with any other particle ($j$), following \eq{eq:ed1}. This continues until either the steady state, or a frozen state with  $x_i>\theta$ for all $i$ is reached. In the case of frozen state, as mentioned earlier,  any one particle is picked up randomly and it loses its energy to any other (randomly chosen) particle, and goes below the threshold $\theta$. This induces further dynamics. Eventually steady state is reached. We study this steady state energy distribution $p(x)$ (see Fig. \ref{fig:mf_dist} ), and the order parameter $O\equiv\int_0^{\theta} p(x) dx$ (see Fig. \ref{fig:mf_order}), showing a ``phase transition'' at $\theta_c\simeq0.607\pm0.001$. A power law fit $O\sim (\theta-\theta_c)^{\beta}$ gives $\beta\simeq 0.97\pm0.01$.  

\begin{figure}[h]
\begin{center}
    \includegraphics[angle=0,width=1.0\linewidth]
        {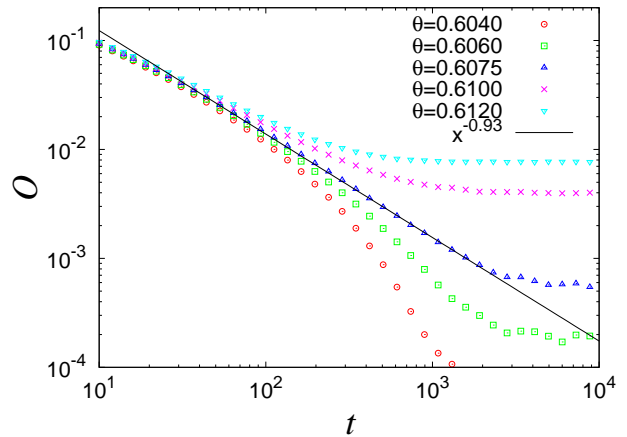}
    \end{center}
\caption{(Color online) Variations of $O$  versus time $t$, shown for different $\theta$ values. At critical value  $\theta_c$, order parameter follows a power law decays  with exponent $\delta\simeq0.93$ (shown by the solid line).  ($N=10^7$; Mean field model with average taken over many independent initial conditions). 
}
\label{fig:relaxation_mean}
\end{figure}

We also studied the relaxation behavior of $O$. At $\theta=\theta_c$, the $O(t)$ variation  fits well with $t^{-\delta}$; $\delta\simeq0.93 \pm 0.01$ (see Fig. \ref{fig:relaxation_mean}). The relaxation time $\tau$ is estimated numerically from the time value at which $O$ first touches the steady state value  $O(\theta)$ within a pre-assigned error limit. We find diverging growth of relaxation time $\tau$  near $\theta=\theta_c$ (see Fig. 4), showing `critical slowing down' at the critical value $\theta_c$. The values of exponent $ z $ for the divergence in $\tau \sim |\theta-\theta_c|^{-z}$ have been estimated  (for both $\theta>\theta_c$ and $\theta<\theta_c$). For the mean-field model, the fitting value for exponent $z \simeq 0.83 \pm 0.01$. 

\begin{figure} [h]
\begin{center}
    \includegraphics[angle=0,width=1.0\linewidth]
        {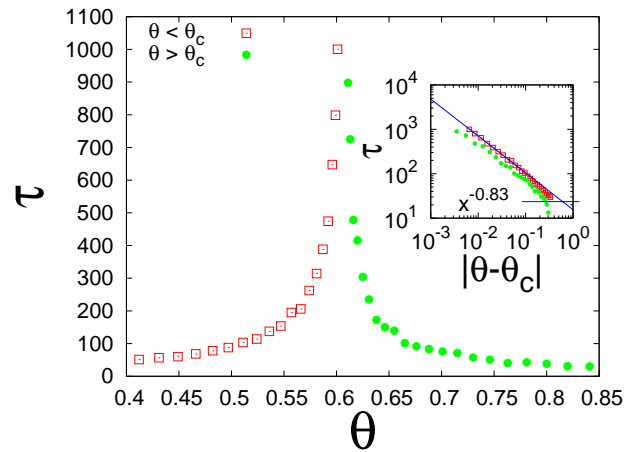}
    \end{center}
\caption{(Color online) Variation of $\tau$ versus  $\theta$. (Inset) Scaling fit $\tau\sim |{\theta-\theta_c}|^{-z}$, with exponent $z \simeq 0.83$ (Mean field case; $N=10^5$).
}
\label{fig:time exp_mean}
\end{figure}

We have also studied the universality of this behavior by generalizing the dynamics in \eq{eq:ed1} to 
\begin{eqnarray}
  {x^<_i}'  &=& \epsilon_1 x^<_i + \epsilon_2 x_j \, ,
  \nonumber \\
 x_j' &=& (1-\epsilon_1) x^<_i + (1-\epsilon_2)x_j \, ,
  \label{eq:ed2}
\end{eqnarray}
where $\epsilon_1$ and $\epsilon_2$  are the random stochastic variables within the range $[0,1]$. The critical point $\theta_c$ shifts to $\theta_c\simeq0.69$ ($\theta_c \simeq0.61$ for $\epsilon_1$= $\epsilon_2$ = $\epsilon$ ). The transition behaviour is seen to be  universal near the critical point $\theta_c$, but the critical point depends specifically on the model (see \fg{fig:mf_univer}).

\begin{figure}
\begin{center}
    \includegraphics[angle=0,width=1.0\linewidth]
        {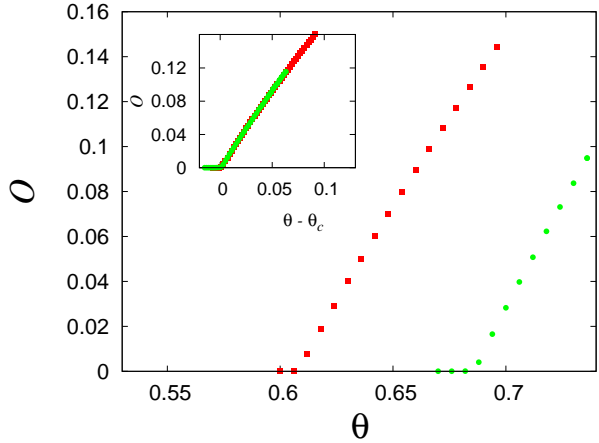}
    \end{center}
\caption{(Color online) Variation of steady state order parameter $O(\theta)$ against $\theta$ for dynamics following \eq{eq:ed1} (denoted by red squares) and \eq{eq:ed2} (denoted by green circles). (Inset) Shows $O$ vs. $(\theta  - \theta_c)$ for both cases. ($N=10^5$; Mean field case).
}
\label{fig:mf_univer}
\end{figure}

\subsection{Results: One dimensional model}
In one dimensional lattice version, the particles  are arranged on a periodic chain. At any time $t$, we randomly select  a  lattice site $i$. If the energy of the corresponding particle is below a prescribed threshold energy $\theta$, then it collides with any one randomly chosen nearest neighbors $j(= i \pm 1)$  which can have any energy whatsoever, and the two particles will exchange energy according to  \eq{eq:ed1}. After each such successful collision, the time is incremented by unity. This process is continued until steady state is reached.  The steady state order parameter $O$ variations against theshold $\theta$ is shown in \fg{fig:1d_order}, with exponent $\beta\simeq0.41\pm0.02$ and $\theta_c\simeq0.810\pm0.001$. The fitting value for exponent $ z $  turn out to be around $ 1.9 \pm 0.05$ (see \fg{fig:1d-time}). Also we find $\delta\simeq0.19\pm0.01$ (see \fg{fig:1d-Relax}). It may be noted that a recent study of a related chain model with such energy cut-off for kinetics, where the effective temperature is varied, has been studied \cite{pk:2011}. Though the behavior is similar, the effective critical behavior (exponent values) seem to be quite  different.

\begin{figure}
\begin{center}
    \includegraphics[angle=0,width=1.0\linewidth]
        {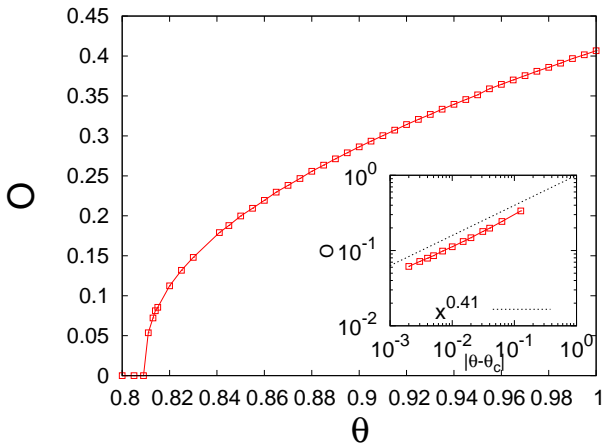}
    \end{center}
\caption{(Color online) Variation of $O$, the average number of •particles  below the threshold energy $\theta$  in the steady state ($t> \tau$), against threshold energy $\theta$, following dynamics of \eq{eq:ed1} for $1$D. ($N=10^4$). (Inset) Shows scaling fit $(\theta-\theta_c)^{\beta}$ with $\beta\simeq 0.41$.
}
\label{fig:1d_order}
\end{figure}

\begin{figure}
\begin{center}
    \includegraphics[angle=0,width=1.0\linewidth]
        {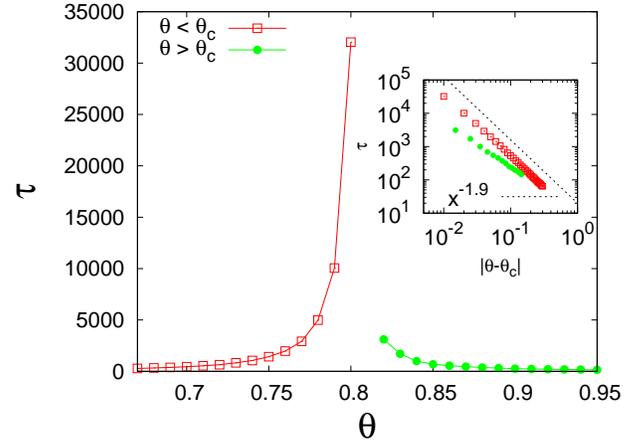}
    \end{center}
\caption{(Color online) Relaxation time $\tau$ as a funtion of $\theta$. Clearly $\tau$ diverges as $\theta\rightarrow \theta_c$. (Inset) Numerical fit to $\tau\sim |{\theta-\theta_c}|^{-z}$, with $z\simeq 1.9 $. ($N=10^4$, 1D case).
}
\label{fig:1d-time}
\end{figure}

\begin{figure}
\begin{center}
    \includegraphics[angle=0,width=1.0\linewidth]
        {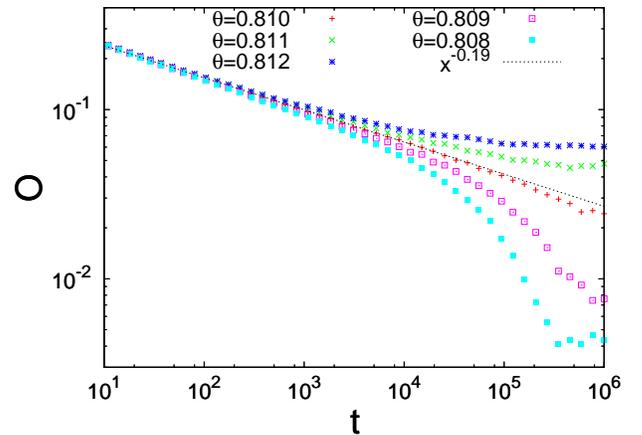}
    \end{center}
\caption{(Color online) Variation of $O(t)$ versus $t$ for different $\theta$ values for 1D case ($N=10^4$). 
}
\label{fig:1d-Relax}
\end{figure}

\subsection{Results: Two dimensional model}

 For the 2D lattice version, the particles  are arranged on a square lattice, and this time one of the four nearest neighbors of $i$ is chosen randomly as particle $j$. 
 If the energy of the particle is below a prescribed threshold energy $\theta$, then it collides with any one randomly chosen nearest neighbor $j$ which can have any energy whatsoever, and the two particles will exchange energy according to  \eq{eq:ed1}. After each such successful collision, the time is incremented by unity. This process is continued until steady state is reached.  Variation of the steady state order parameter $O$  against theshold $\theta$ is shown in \fg{fig:2d-order}, with exponent $\beta\simeq0.67\pm0.01$ and $\theta_c\simeq0.675\pm0.005$. 
 Also, we find $z\simeq1.2\pm0.01$ (see \fg{fig:2d-time}) and $\delta\simeq0.43\pm0.02$ (see \fg{fig:2d-relax}).

\begin{table}[h]

\begin{tabular}{|l|l|c||c|}
\hline
 & & This Model & Manna Model  \\ \hline
\multirow{3}{*}{$\beta$} & 1D  & 0.41 $\pm$ 0.02 & 0.382 $\pm$ 0.019 \\
 & 2D & 0.67 $\pm$ 0.01  & 0.639 $\pm$ 0.009  \\
 & MF & 0.97 $\pm$ 0.01 & 1  \\
 \hline
\multirow{3}{*}{$z$} & 1D  & 1.9 $\pm$ 0.05 & 1.876 $\pm$ 0.135 \\
 & 2D  & 1.2 $\pm$ 0.01 & 1.22 $\pm$ 0.029 \\
 & MF & 0.83 $\pm$ 0.01 & 1  \\
\hline
\multirow{3}{*}{$\delta$} & 1D  & 0.19 $\pm$ 0.01 & 0.141 $\pm$ 0.024 \\
 & 2D & 0.43 $\pm$ 0.02 & 0.419 $\pm$ 0.015  \\
 & MF & 0.93 $\pm$ 0.01 & 1  \\
\hline
\end{tabular}
\caption{Comparison of critical exponents of this model with those of the Manna model \cite{Lubeck:2004}.}\label{table:exponents}
\end{table}

\begin{figure}
\begin{center}
    \includegraphics[angle=0,width=1.0\linewidth]
        {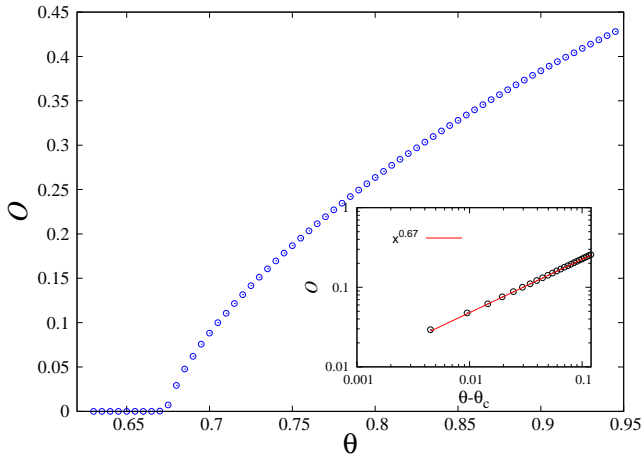}
    \end{center}
\caption{(Color online) Variation of $O$, the average number of particles  below the threshold energy $\theta$  in the steady state ($t> \tau$), against threshold energy $\theta$, following dynamics of \eq{eq:ed1} for $2D$ case. The simulation is done for lattice size $1000$ $\times$ $1000$. (Inset) Shows scaling fit $(\theta-\theta_c)^{\beta}$ with $\beta\simeq 0.67$.
}
\label{fig:2d-order}
\end{figure}

\begin{figure}
\begin{center}
    \includegraphics[angle=0,width=1.0\linewidth]
        {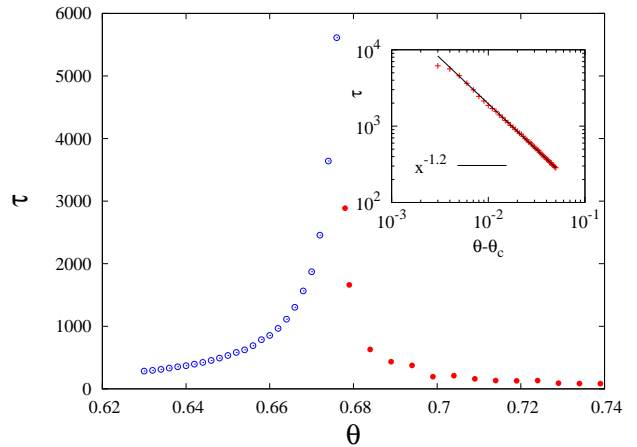}
    \end{center}
\caption{(Color online) Relaxation time  ($\tau$) diverges as $\theta$ approches $\theta_c$ from both sides. (Inset) Numerical fit to $\tau\sim |{\theta-\theta_c}|^{-z}$ for $\theta<\theta_c$ ( $z\simeq1.2\pm 0.01$).  (Simulations for $100$ $\times$ $100$ system, 2D case).
}
\label{fig:2d-time}
\end{figure}

\begin{figure}
\begin{center}
    \includegraphics[angle=0,width=1.0\linewidth]
        {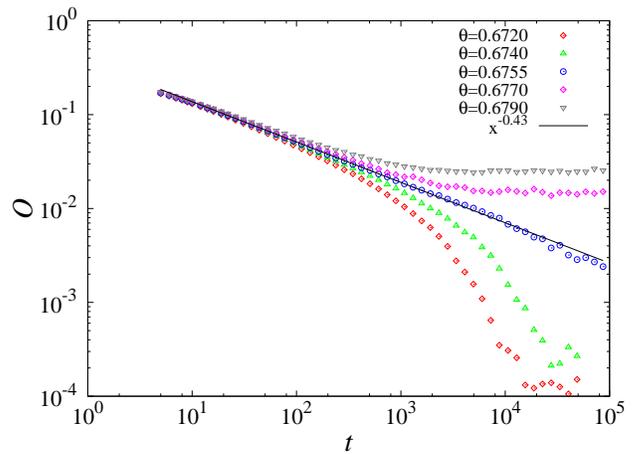}
    \end{center}
\caption{(Color online) Variation of $O(t)$ versus time $t$, shown for different $\theta$ values for $2D$ case. At critical value $\theta_c$, order parameter follows a power law decay  with exponent $\delta\simeq0.43$. The simulation is done for lattice size $500$ $\times$ $500$.
}
\label{fig:2d-relax}
\end{figure}


All these estimated values of the critical exponents $\beta$, $z$, and $\delta$ are summarized in Table \ref{table:exponents}.


\section{Finite Size Effect}

\begin{figure*}[t]
\includegraphics[width=5.9cm]{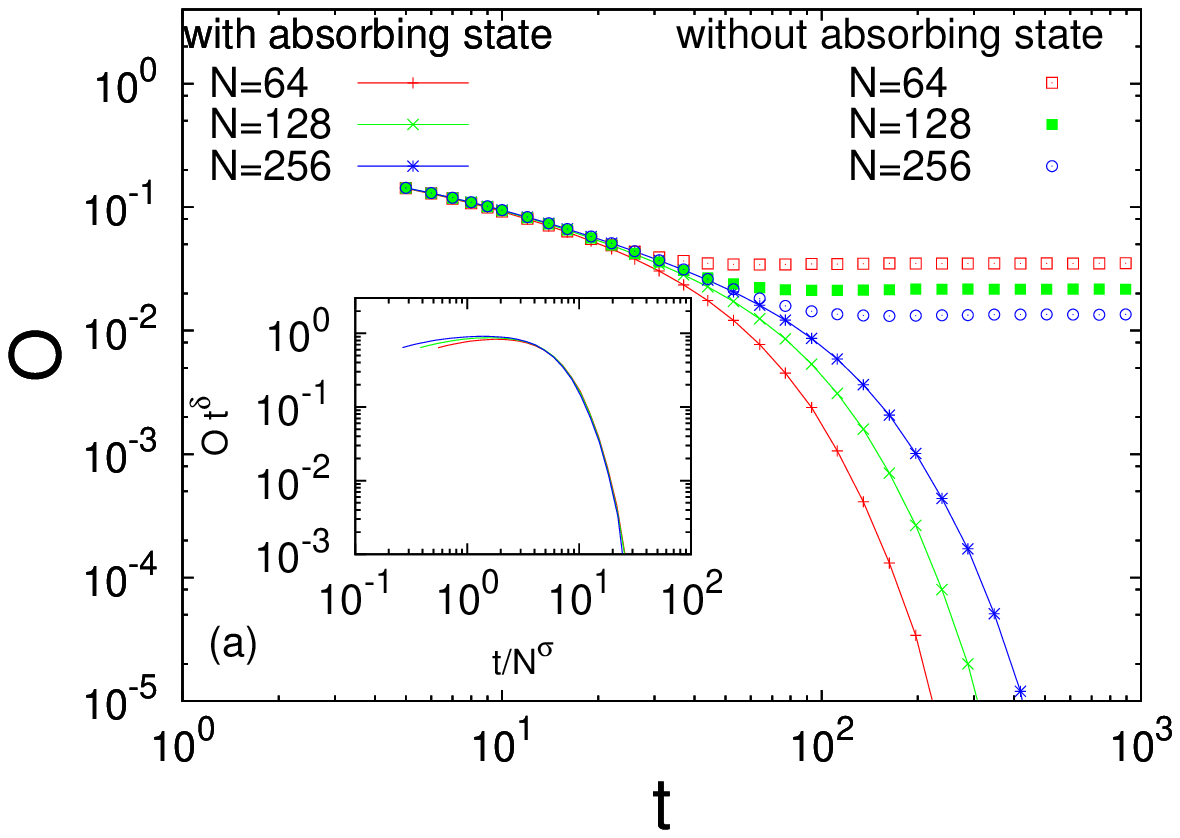}
\includegraphics[width=5.9cm]{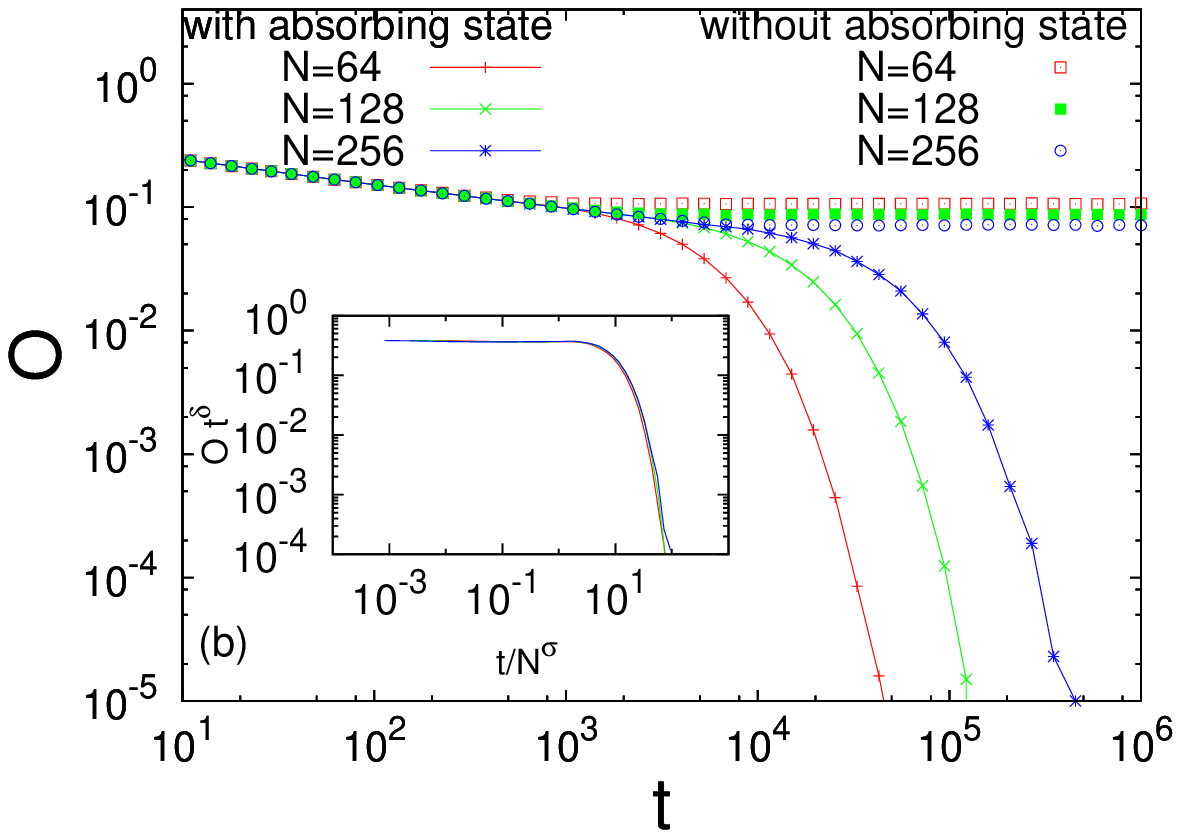}
\includegraphics[width=5.9cm]{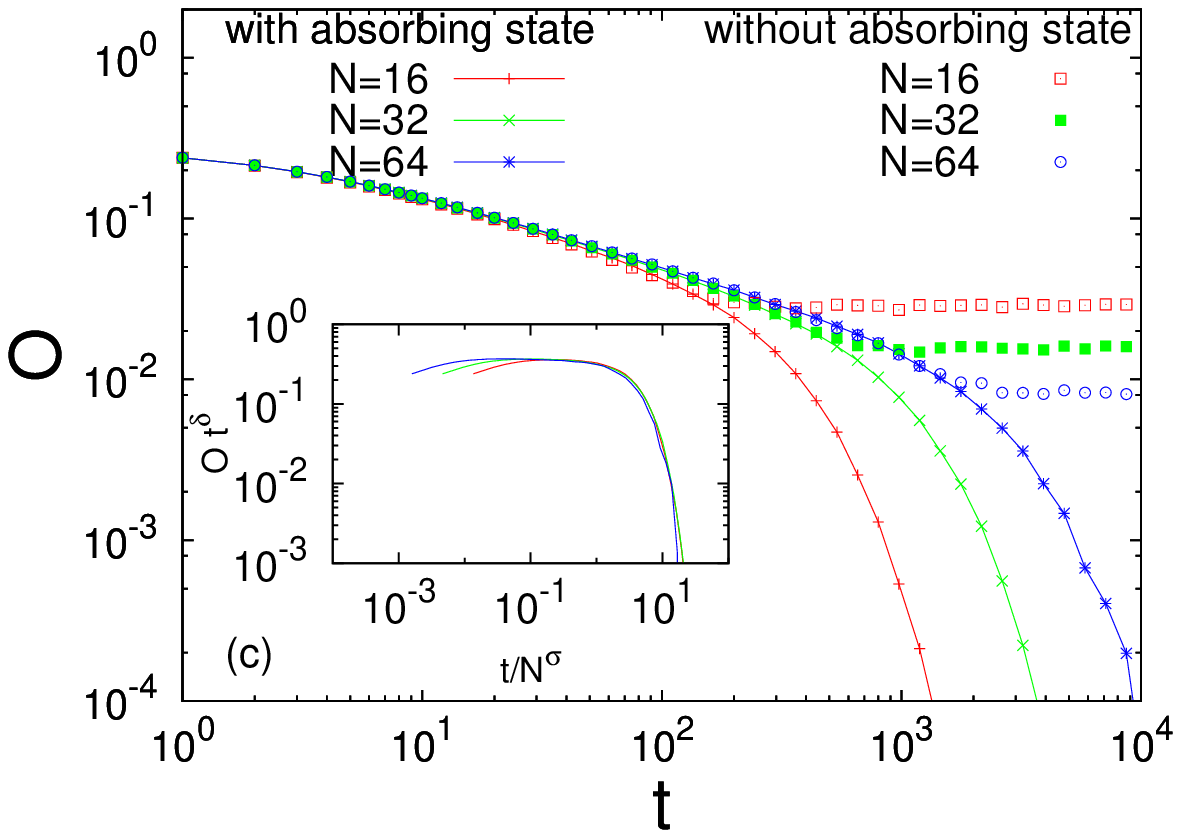}
\caption{Study of finite size effect:
The variation of $O(t)$, versus $t$ at $\theta=\theta_c$ for systems of different sizes $N$ are shown. 
The plots with bare symbols correspond to the case when there is no absorbing state and those with symbols connected by solid lines correspond to the presence of absorbing state.
The Figs. (a), (b) and (c) correspond to MF, 1D and 2D respectively. 
Inset: The plot of $O t^{\delta}$ vs. $t/N^{\sigma}$ in presence of absorbing state, for different system size $N$ collapse onto a single curve for $\delta=0.93, 0.19,$ and $0.43$ and $\sigma=0.53, 1.53,$ and  $1.55$ for MF, 1D, and 2D respectively. }
\label{fig:size-effect}
\end{figure*}

We have also studied the time variation of $O$ at different sizes ($N$) at $\theta_c$. Plots of $O(t)t^\delta$ as a function of $t/N^{\sigma}$ for different values of the 
system size $N$ (at critical point) are expected to collapse on a single curve. However, as we have used a special dynamics which never allows the system to fall in the absorbing state, in this case the activity saturates at a small steady value (see Fig. \ref{fig:size-effect}) instead of showing the finite size cut-off.

To study finite size effects in the decay of activity at the critical point, one has to remove the perturbation and allow the system to be trapped in the absorbing states. Using this dynamics, we have studied the effects of finite system size. Fig. \ref{fig:size-effect}, show the decay of $O(t)$ with $t$ at the critical point for MF, $1D$, and $2D$  systems, respectively. The insets in the corresponding figures show the data collapse. The  fitting values of the   exponent $\sigma$ are $\sigma= 0.53\pm 0.02, 1.53 \pm 0.05$ and $1.55 \pm 0.05$ for MF, $1D$, and $2D$ respectively, with $\delta$ values given in Table-I.

\section{Summary \& Discussion}
\label{sec:summary}
 
Inspired by the success of the kinetic exchange models of market dynamics (see e.g., \cite{SCCCbook2010, Yakovenko2009, Arnab2007}) and the observation that the poor or economically backwards in the society take major initiative in the market dynamics (see e.g., \cite{Iglesias-physica, Iglesias-sc&cul} and also \cite{Manna2010, Bak}), we consider an ideal-gas-like model of gas (or market) where at least one of  the particles (or agents) has energy (money) $x$ less than a threshold (poverty line) value $\theta$ takes the initiative to scatter (trade) with any other particle (agent) in the system, following energy (money) conserving random processes (following \eq{eq:ed1}). For $\theta\rightarrow\infty$, the model reduces to the kinetic model of ideal gas with Gibbs distribution. The steady state is found to be ergodic (steady state results are independent of initial conditions). The perturbation employed in the frozen cases ($x_i>\theta$ for all $i$) also does not affect significantly the thermodynamic quantities (e.g., the steady state value of $O$ for $\theta<\theta_c$ goes to 0 with $1/N$, as can be seen from inset (a) of \fg{fig:mf_order}).

In general, we find that the steady state distribution $p(x)$ (see \fg{fig:mf_dist} for the mean field, where each particle can interact irrespective of their distance from the active particle or agent having energy or money less than threshold or poverty line $\theta$) differs in form significantly from the Gibb's distribution, for finite values of $\theta$. The order parameter $O$, giving the average fraction of particles (or agents) having energy (or money) below $\theta$ shows a phase transition behavior: $O=0$ for $\theta<\theta_c$ and $O\ne 0$ for $\theta>\theta_c$ (see \fg{fig:mf_order} for mean field, \fg{fig:1d_order} for 1D, and \fg{fig:2d-order} for 2D, respectively). The critical values are given by $\theta_c\simeq0.61, 0.81, 0.68$ for mean field, 1D and 2D cases, respectively. The variation of $O$ near $\theta_c$ is quite universal (see \fg{fig:mf_univer}). We find $O\sim(\theta-\theta_c)^{\beta}$ with $\beta\simeq0.97, 0.41,0.67$ for mean field, 1D and 2D cases, respectively. We also find that the relaxation time $\tau$ diverges strongly near $\theta_c$ as $\tau\sim(\theta-\theta_c)^{-z}$ with $z\simeq 0.83, 1.9,1.2 $ for mean field, 1D and 2D cases, respectively. Finally, at $\theta=\theta_c$, $O(t)\sim t^{-\delta}$ where $\delta\simeq 0.93, 0.19, 0.43$ for mean field, 1D and 2D cases, respectively (see Table \ref{table:exponents}). 

It might be mentioned here that the above exponent values are indeed very close to those of the Manna Universality (MU) class (\cite{Lubeck:2004}; see also \cite{Manna,Menon}) in mean field, 1D, 2D cases: Our estimates for $\beta\simeq0.97$, $0.41$ and $0.67$ for mean field, 1D and 2D cases, respectively, are quite close to $\beta \simeq 1$, $0.38$ and $0.64$ for corresponding MU cases; $\delta\simeq0.93$, $0.19$ and $0.43$ for mean field, 1D and 2D cases, are also close to $\delta\simeq1$, $0.14$ and $0.42$ in the corresponding MU cases. However, it may be noted that  significant differences in the above estimates do exist. Also, $z\simeq 0.83, 1.9, 1.2 $ for mean field, 1D and 2D cases might be compared with $z\simeq 1, 1.87, 1.22 $ in the corresponding MU cases. These discrepancies could be due to finite size effect, and in that case the critical behavior of our model would belong to the MU class.  As one can see, the estimated values of  the exponents  $\beta$, $\delta$ and $z$ fit reasonably with the scaling relation $\delta=\beta/z$ within our limits of accuracy. In this connection, it is worth mentioning that   the violation of the above scaling relation has also been observed \cite{Rossi}, though  such discrepancies seem to get removed if one uses all sample averages instead of averages over surviving samples \cite{Lee}. In our case, however, this scaling relation seems to hold, as our simulation results correspond to all sample averages.


In summary, when the energy threshold $\theta$ is introduced in the kinetic theory of ideal gas such that the stochastic energy conserving scatterings between any two particles can take place only when one has energy less than $\theta$, the gas system shows an intriguing dynamic phase transition at $\theta=\theta_c$, having the exponent values in the mean field (long range scattering exchange), one dimension and two dimension,   as estimated here using Monte Carlo simulation, are given in Table I.

\begin{acknowledgments}
We are grateful to  Arnab Chatterjee and Gautam Menon for useful comments on the critical behaviour of the model.
We also acknowledge discussions with Mahashweta Basu, Rakesh Chatterjee and Soumyajyoti Biswas.
\end{acknowledgments}


\begin{thebibliography}{}

\bibitem{CCCbook2006} (Eds.) B.K. Chakrabarti, A. Chakraborti and A. Chatterjee, \textit{Econophysics and Sociophysics: Trends and Perspectives} (Wiley-VCH, Berlin, 2006).

\bibitem{SCCCbook2010} S. Sinha, A. Chatterjee, A. Chakraborti and B.K. Chakrabarti, \textit{Econophysics: An Introduction} (Wiley-VCH, Berlin, 2010).


\bibitem{Yakovenko2009}V.~Yakovenko, J.~Rosser, \textit{Rev. Mod. Phys.} \textbf{81}, 1703 (2009).

\bibitem{Arnab2007}A.~Chatterjee, B.K.~Chakrabarti, \textit{Eur. Phys. J.} \textbf{B {60}}, 135 (2007).

\bibitem{Patriarca2010}M.~Patriarca, E.~Heinsalu, A.~Chakraborti, \textit{Eur. Phys. J.} \textbf{B {73}}, 145 (2010).

\bibitem{Chakraborti2010a}A. Chakraborti, I. Muni Toke, M. Patriarca, F. Abergel, \textit{Quantitative Finance}, in press; available at \textit{arXiv:0909.1974v2} (2010).

\bibitem{Redner2010}P. L. Krapivsky, S. Redner, \textit{Science and Culture (Kolkata)} \textbf{76}, 424 (2010); \textit{arXiv:1006.4595}  [physics.soc-ph].

\bibitem{Chakraborti2000a}A.~Chakraborti, B.K. Chakrabarti, \textit{Eur. Phys. J.} \textbf{B {17}}, 167 (2000).

%
%
\bibitem{Chatterjee2004a}A.~Chatterjee, B.K. Chakrabarti, S.~S. Manna, \textit{Physica} \textbf{A {335}}, 155 (2004).

%
\bibitem{Chakraborti2009}A.~Chakraborti, M.~Patriarca, \textit{Phys. Rev. Lett.} \textbf{{103}}, 228701 (2009).

\bibitem{Mehdi2010}M. Lallouache, A. Chakraborti and B.K. Chakrabarti, \textit{Science and Culture (Kolkata)} \textbf{76}, 485 (2010); \textit{arXiv:1006.5921}  [physics.soc-ph].

\bibitem{LCCC2010}M. Lallouache, A.S. Chakrabarti, A. Chakraborti and B.K. Chakrabarti, \textit{Phys. Rev.} \textbf{E {82}}, 056112 (2010).

\bibitem{CC2010}A. Chakraborti and B.K. Chakrabarti, in Eds. F. Abergel, B.K. Chakrabarti, A. Chakraborti and M. Mitra, \textit{Econophysics of order-driven markets} (Springer- Verlag (Italia), Milan, 2011).

\bibitem{Sen} P. Sen, \textit{Phys. Rev.} \textbf{E {83}}, 016108 (2011).

\bibitem{Iglesias-physica} S. Pianegonda, J. R. Iglesias, G. Abramson and J. L. Vega, Physica A \textbf{322}, 667 (2003).

\bibitem{Iglesias-sc&cul}  J.R. Iglesias, \textit{Science and Culture (Kolkata)} \textbf{76 }, 437 (2010).

\bibitem{pk:2011}
M. Basu, U. Gayen and P. K. Mohanty, arXiv.1102.1631v1.

\bibitem{Lubeck:2004}
S. L\"{u}beck, \textit{Int. J. Mod. Phys. B} \textbf{18}, 3977 (2004).


\bibitem{Manna2010}
A. Chakraborty, S. S. Manna \textit{Phys. Rev.}  \textbf{E {81}}, 016111 (2010).

\bibitem{Bak}
P. Bak and K. Sneppen, \textit{Phys. Rev. Lett.}  \textbf{71}, 4083 (1993).






\bibitem{Manna}S. S. Manna, \textit{Physica A} \textbf{179}, 249(1991); S. S. Manna, \textit{J. Phys. A }\textbf{24}, L363 (1991).

\bibitem{Menon}
G. I. Menon, S. Ramaswamy,  \textit{Phys. Rev.} \textbf{E {79}}, 061108 (2009).


\bibitem{Rossi}
M. Rossi, R. Pastor-Satorras, and A. Vespignani \textit{Phys. Rev. Lett.} \textbf{85}, 1803 (2000).

\bibitem{Lee}
S. B. Lee and S. G. Lee \textit{Phys. Rev.} \textbf{E {78}}, 040103(R) (2008).


\end{thebibliography}
\end{document}